\documentclass[conference]{IEEEtran}
\IEEEoverridecommandlockouts
\usepackage{graphicx}
\usepackage{color}
\usepackage{soul}

\begin{document}

\title{Coconut Trees Detection and Segmentation in Aerial Imagery using Mask R-CNN}


\author{\IEEEauthorblockN{Muhammad Shakaib Iqbal\IEEEauthorrefmark{1}, Hazrat Ali \IEEEauthorrefmark{1},
Son N. Tran\IEEEauthorrefmark{2}, 
and Talha Iqbal}\IEEEauthorrefmark{3}

\IEEEauthorblockA{\IEEEauthorrefmark{1}Department of Electrical and Computer Engineering,\\ COMSATS University Islamabad, Abbottabad Campus, Abbottabad, Pakistan. \\Email: hazratali@cuiatd.edu.pk}
\IEEEauthorblockA{\IEEEauthorrefmark{2}Department of Information and Communication Technology, University of Tasmania, Australia.}
\IEEEauthorblockA{\IEEEauthorrefmark{3}Smart Sensor Lab, School of Medicine, National University of Ireland, Galway, Ireland.
}
\thanks{*Correspondence to: hazratali@cuiatd.edu.pk}
\thanks{This work is published in IET Computer Vision. doi: 10.1049/cvi2.12028}
}

\maketitle
\begin{abstract}
Food resources face severe damages under extraordinary situations of catastrophes such as earthquakes, cyclones, and tsunamis. Under such scenarios, speedy assessment of food resources from agricultural land is critical as it supports aid activity in the disaster-hit areas. In this work, we present a deep learning approach for the detection and segmentation of coconut trees in aerial imagery provided through the AI competition and organized by The World Bank in collaboration with \emph{OpenAerialMap} and \emph{WeRobotics}. We use Masked Region-based Convolution Neural Network (Mask R-CNN) approach for coconut trees identification and segmentation. For the segmentation task, we use Mask R-CNN model with ResNet50 and ResNet101 based architectures. We perform several experiments with different configuration parameters and report the best configuration for the detection of coconut trees with more than 90\% confidence factor. For evaluation purpose, we use the Microsoft COCO dataset evaluation metric namely mean average precision (mAP). We achieve an overall 91\% mean average precision for coconut trees detection.
\end{abstract}

\maketitle

\section{Introduction}\label{sec1}
Natural disasters in the Kingdom of Tonga (South Pacific) are an unfortunate global reality. Their consequences can be damaging for the population of the south pacific who heavily depend on the local agriculture as a primary food source \cite{reff1}. As per the 2015 statement of the Secretary-General of the UN on the ``Implementation of the International Strategy for Disaster Reduction'',  approximately 1.5 trillion USD losses have incurred as a direct consequence of the natural catastrophes around the world \footnote{https://www.unisdr.org/}. The rate of recurrence as well as the magnitude of the severity of these disasters are increasing. Hence, there is a great demand to reinforce food security mechanisms and make appropriate assessments of the damages caused \cite{reff2}.

When cyclones strike, recognizing the area of damage is crucial for effective humanitarian response and securing undamaged food sources like coconut trees. The World Bank seeks qualified teams to develop machine learning based methods to automate the assessment of aerial imagery and to classify and locate the standing trees such as coconut trees within the aerial snapshot \cite{reff3}. Manual aerial image classification is a resource as well as skill$-$intensive task and requires a lot of time. More importantly, manual aerial image classification is not typically risk-free in disaster-hit regions.

OpenAerialMap, World Bank, and WeRobotics have collaboratively launched an open machine learning challenge to speed up the classification and analysis of high-resolution aerial imagery before and after humanitarian disaster \cite{reff4}. The idea is to explore and develop machine learning solutions for the classification of various features of interest in aerial imagery obtained through UAV. The features thus obtained can then be utilized for object detection and classification to help in the assessment of damages caused. One of the tasks in the challenge is to build a model for coconut trees detection. In this task, we are given a spatial high-resolution image (about eight cm/pixel), which covers 50 $km^2$ Area of Interest (AOI) of the kingdom of Tonga in the south pacific region. The imagery is taken during October 2017, which is quite recent. Along with the aerial image, we are given shape (.shp) files, to recognize the geometric locations and classes of the targets (objects of interest) like roads and trees. Data for relevant features have been labeled by the group of volunteers from the humanitarian OpenStreetMap (OSM) community\footnote{https://www.openstreetmap.org}.

Object detection in aerial imagery is an interesting task and has attracted the computer vision and machine learning research community \cite{reff5yang2018deep}, \cite{reff6sommer2017fast}, \cite{reff7luque2017spatio}. Typically, these approaches use Convolutional Neural Networks (CNN) for object detection. The prevailing work has mostly been done on road detection and vehicle detection  \cite{reff5yang2018deep}, \cite{reff6sommer2017fast}, \cite{reff7luque2017spatio}. In this work, we choose to develop a framework to detect and locate coconut trees. More specifically, it addresses the task of coconut trees classification and localization. With the help of experimental results, we demonstrate how we can use mask R-CNN to detect coconut trees within the images. This is challenging as some of these images include mislabeled and missing ground truth entries. Besides, generic shapes that have different objects are difficult to be differentiated in the aerial image. Finally, there are many small objects occupied by densely concentrated regions in the aerial image. 

This contributions of this paper are as follows:
\begin{itemize}
    \item We present a framework for automatic detection and localization of coconut trees within given aerial imagery. The framework is able to detect each individual coconut tree with a high confidence factor and provide a segmented mask. 
  \item The proposed framework for coconut trees detection provides a baseline approach, which can be easily extended to detect and identify other types of trees.
  \item Agriculture resource management is a labor-intense and high-risk job. The proposed approach provides a low-cost solution for agriculture resource management and measurement of the impact of disasters on natural food resources while reducing the risk factors for human operators. 
\end{itemize}

The rest of the paper is organized as follows: In Section  \ref{sec:RelatedWork} we provide a brief overview of deep learning techniques for objection detection. In Section \ref{sec:SystemModel}, we provide a detailed description of the dataset, methods, and training mechanism used in this work. We present the results on coconut trees prediction with the help of figures in Section \ref{sec:results}, and also provide a discussion on the model and results obtained. Finally, we conclude our work in Section \ref{sec:Conclusion}.

\section{Related Work}
\label{sec:RelatedWork}
Over the last tens of years, satellite imagery has been often used in a diversified range of applications ranging from forestry \cite{reff8dalponte2008fusion} to agriculture \cite{reff9frolking2002combining},\cite{reff10rhee2010monitoring}, target detection \cite{reff11li2010saliency} and regional planning to warfare \cite{reff12campbell2011introduction}. Satellite imagery has also been broadly employed to monitor natural disasters and various other adverse incidents to investigate their impact on the environment.

Deep learning modifies the traditional machine learning by addition of more ``depth'' in the model and transforming the information through several layers and non-linearity functions. This provides hierarchical data representation through abstraction of many levels \cite{reff13schmidhuber2015deep}. Deep learning extracts useful features from raw data, with features from high levels of the hierarchy shaped through a combination of low level features \cite{reff14rusk2015deep}. The huge parallelization possible in deep learning models enables us to develop highly complex models for learning complex features and performing extremely well on many AI tasks \cite{reff15pan2009survey}. So, deep learning models can enhance categorization efficiency or minimize error in regression problems, given adequate large data is available for a specific domain task. 

The large capacity of models and highly hierarchical structure performs very well particularly on the prediction and classification tasks, being adaptable and flexible for a broad range of highly complex problems \cite{reff15pan2009survey}. Although deep learning has got fame in various applications coping with raster-based information (such as pictures, video), it can be applied to an array of different types of information, i.e. speech, audio and natural language, and other data types like population information \cite{reff16demmers2012simultaneous}, continuous data such as weather data \cite{reff17sehgal2017crop} and soil chemistry \cite{reff17sehgal2017crop}. The vital role of utilizing deep learning in the processing of images is the reduced need for feature engineering. In the past, conventional methods for image classification tasks were typically based on manual hand-engineered features. However, features engineering is a time-consuming, costly, and complex method that needs to be changed whenever the data-set or the problem changes. Thus, feature engineering involves a costly effort, which is based on the expert's ability and may not generalizes well \cite{reff18amara2017deep}. Alternatively, a deep learning model does not rely on feature engineering and rather learn features through representation learning.  

The region-based convolutional neural network (R-CNN) has proved to be very successful for segmentation tasks \cite{reff20zhao2019object}, \cite{reff21girshick2015region}. In R-CNN, a selective search technique is applied to detect region proposals within the input image. Region proposals structure the features vector that is given to multiple classifiers to represent a distribution of class variables and also to a regression model to refine bounding boxes of regions of a proposal. Fast R-CNN and Faster R-CNN proposed in \cite{reff22girshick2015fast}, \cite{reff23ren2015faster} respectively, accelerate the detection procedure by first applying a deep CNN to the input image and then extracting features map and simply swapping selective-search by Region Proposal Network (RPN) to create region proposals, predicting bounding boxes and classes of the objects. The extension of Faster R-CNN to Mask R-CNN \cite{reff24he2017mask} puts a parallel branch to object detection to predict object masks with very small overhead. Mask R-CNN outperformed top models in the 2017 COCO competition in segmentation, object detection, and bounding-box detection. 

Hence, we prefer Mask R-CNN as our model. Recently reported attempts on object detection and image detection include \cite{maghsoudi2020automatic} and \cite{liang2020polytransform}. The work in \cite{maghsoudi2020automatic} uses the U-Net architecture with a ResNet decoder and the work in \cite{liang2020polytransform} produces geometry preserved masking for a better fit on the object's boundaries. The selection of Mask R-CNN in our work was made based upon two considerations 1) The model outperformed top models on the most recent COCO competition for object detection and 2) the underlying software tools and tensor-flow libraries make the import of the pre-trained weights relatively easier making the implementation more convenient. 

\section{Proposed Methodology}
\label{sec:SystemModel}
The overall pipeline is summarized in Figure 1. Briefly, we pre-process the data and divide the data into training and test sets. Since we utilize pre-trained weights, it is not necessary to train the whole neural network. We train the final layers (bounding box heads/classification) and select the configuration settings with the minimum validation error. At last, we evaluate the overall performance using the previously unseen test data. In the following sub-sections, we talk about all these phases in more detail. To generalize our model, we have taken the following steps:
\begin{itemize}
\item We use a stronger weight decay i.e., the L2 regularization.
\item We use an optimal learning rate of 0.001 (to avoid converging to local minima) after trying out different values among \{0.0001, 0.0003, 0.001, 0.003, 0.01\}.
\item We use an adaptive optimizer named as SWATS, an approach to switch optimization from Adam to Stochastic gradient descent and thus achieving better generalization \cite{richardsocher2017}.
\end{itemize}

\begin{figure}[tp!]
\centering
\graphicspath{{./figures/}}
\includegraphics[width=\columnwidth]{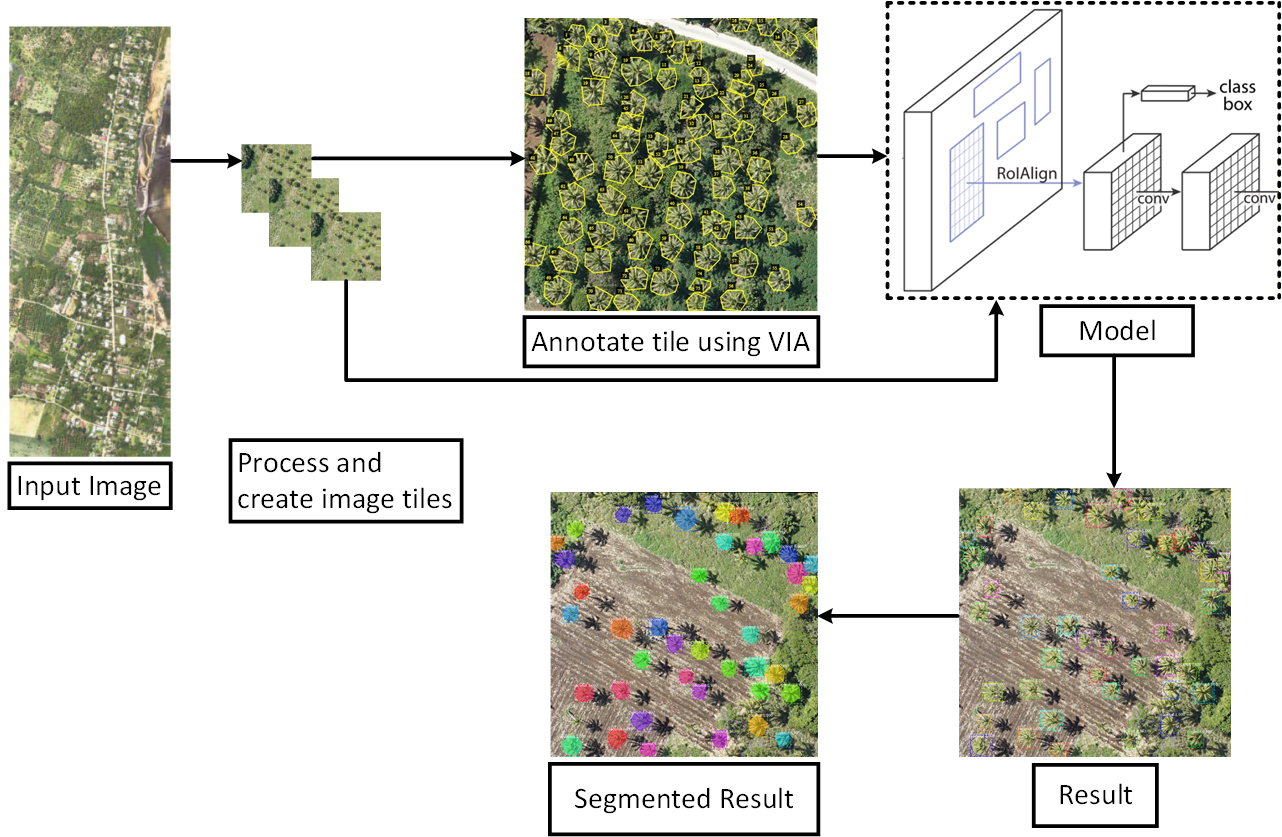}
\caption{Proposed Model Pipeline (From left to right) Original snapshot is divided into tiles divided into training and test examples. Annotated images are fed to the pre-trained model. Final layers of the model are fine-tuned on our own dataset after that model is ready for detection.}
\label{fig:System}
\end{figure}
\begin{figure}[!ht]
\centering
\graphicspath{{./figures/}}
\includegraphics[width=\columnwidth]{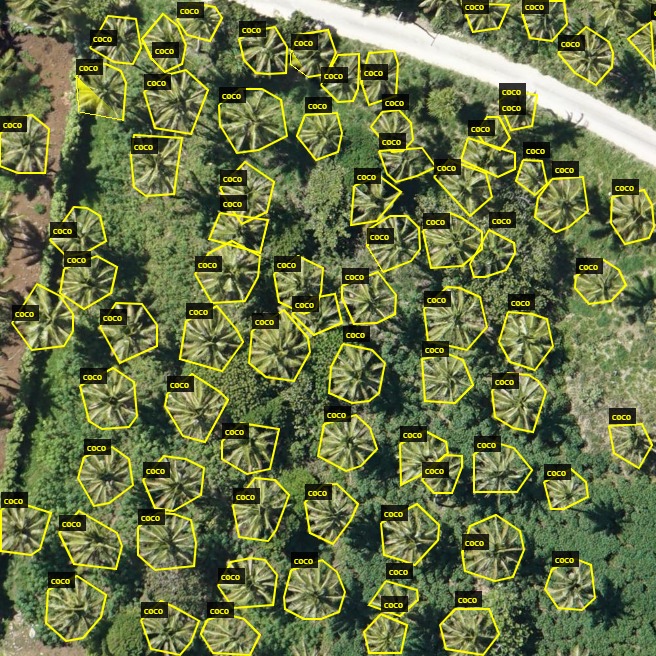}
\caption{Sample of training data. The image annotation is done using VGG image annotator. Only coconut trees are annotated.}
\label{fig:Network}
\end{figure}
\subsection{Data Processing}
As we discussed in the introduction section, we are provided with one single high-resolution aerial snapshot and a shape file (GIS file that stores geometric locations as well as attributes of geometric features, such as points, lines, and polygons). Our objective is to combine and convert both data sources into training, validation, and test sets appropriate for our object detection model. GDAL is an open-source library to deal with raster and vector geospatial data types\footnote{https://gdal.org/}. Fiona is a very popular python library for writing and reading geospatial records\footnote{https://pypi.org/project/Fiona/}. We use Fiona to read the shape file into a JSON file format where we are given the geographical data and it is readily available for data processing. After that, we extract the positions of the items of interests, such as for example coconut trees by looking at the tags offered for every geo JSON object in the shape file. The positions are given in the latitudinal and longitudinal coordinate system. To be able to map these locations on to the very high-resolution aerial image, we convert the actual image into the latitude-longitude coordinate system by using GDAL tools. We furthermore map the latitude and longitude coordinates taken out from the shape file into the image-pixels by using the geographical metadata in the high-resolution aerial image. At the end of the procedure, we have an image combined with the pixels where objects of interests (coconuts trees) are located.

The input to the Mask R-CNN framework is the set of annotated train image tiles of size $1000\times1000$ pixels. We subdivide the actual image into patches of the dimensions $1000\times1000$. We take 70 such tiles and manually annotate every single tile by positioning the coconut trees and then drawing polygons around them. It really is a time-consuming procedure, but we believe correctly annotated dataset is vital for training a model with high prediction accuracy. We use VGG Image Annotator\footnote{http://www.robots.ox.ac.uk/~vgg/software/via/} to promptly annotate 70 image tiles. Every single annotation consists of JSON file format and keeps the positions of all polygons along with their tags. We now have approximately between 40 to 60 objects in each tile, therefore we conclude that the training dataset is substantial. A good example of an image tile annotated with the VGG Image Annotator is displayed in Figure 2. 

It is very important to note, that during the annotation we find a lot of discrepancies. For example, quite a few coconut trees were mislabeled (such as coconut trees labelled as bananas, dark areas and shadows labeled as trees) and several trees were not marked at all.

\subsection{Deep Learning Architecture and Training}
As we discussed in the model selection phase, the foundation of our approach is based on the Mask R-CNN implementation. The ResNet101, a deep residual neural network with 101 layers, is the backbone architecture that extracts feature maps from the input image \cite{reff27he2016deep}. Residual networks enable us to efficiently train deep neural networks simply by introducing skip connections, in which weights coming from previous layers are copied into a more deep layer. It requires an image of $1000\times1000\times3$ and then outputs feature map of dimension $32\times32\times2048$. These features are moved to an RPN for training regression/classification of object classes and generation of bounding boxes.
\begin{figure}[!ht]
\centering
\graphicspath{{./figures/}}
\includegraphics[width=\columnwidth,height=4cm]{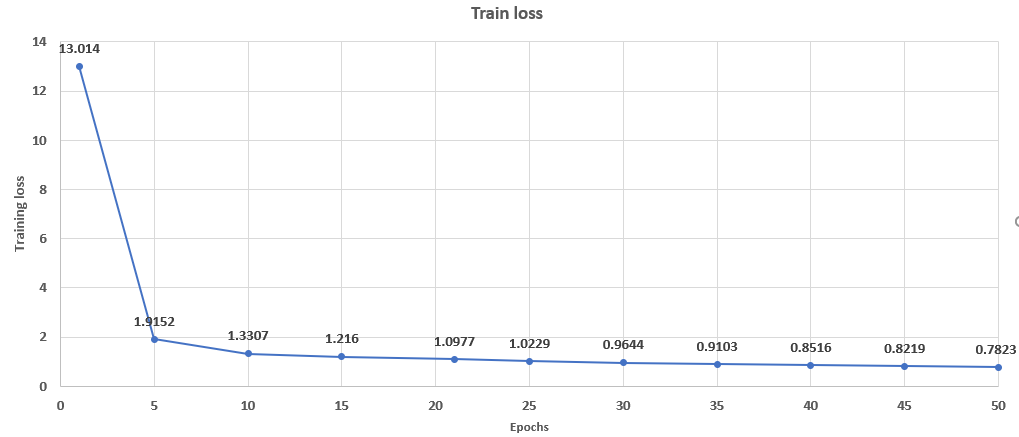}
\caption{Training loss from epoch 1 to 50. Changes in the training loss after 21 epochs are not significant.}
\label{fig:train_loss}
\end{figure}
We initiated the training procedure by downloading a model pre-trained on the Microsoft COCO dataset, one of the most widely used datasets for object detection and segmentation. In principle, we do not change the setting of earlier layers but modify few RPN parameters as we aim to train the final layers of the model (referred to as regression and classification heads). To accelerate the training process, although we aim to attain high accuracy, we specify the minimum region proposal confidence to 0.9, which means only regions with more than 90\% confidence of potentially containing trees are considered. The confidence score of 0.9 is selected after an empirical evaluation of different possible values. Selecting a value less than 0.9 causes classifiers to incorrectly detect shadows and other trees like objects in the image as coconut trees (more false positives), as shown in Figure \ref{fig:Con_0.7} . On the other, a confidence value greater than 0.9 results in missing out some of the coconut trees, such as trees behind light clouds, are not detected properly (more false negatives), eventually resulting in a lower detection accuracy, as shown in Figure \ref{fig:Con_0.95}. Furthermore, since in RGB aerial images coconut trees are expected to have approximately similar aspect ratios and sizes, anchor scales are set between 10 to 130. The learning rate is set to 0.001 while using the weight decay of 0.0001. We divided the available data (70 tiles) into training/validation/test sets of 50/10/10 image tiles and perform several experiments by changing the number of steps, the number of epochs, and the number of maximum possible Regions of Interest (ROIs).
\begin{figure}[!ht]
\centering
\graphicspath{{./figures/}}
\includegraphics[width=\columnwidth,height=4cm]{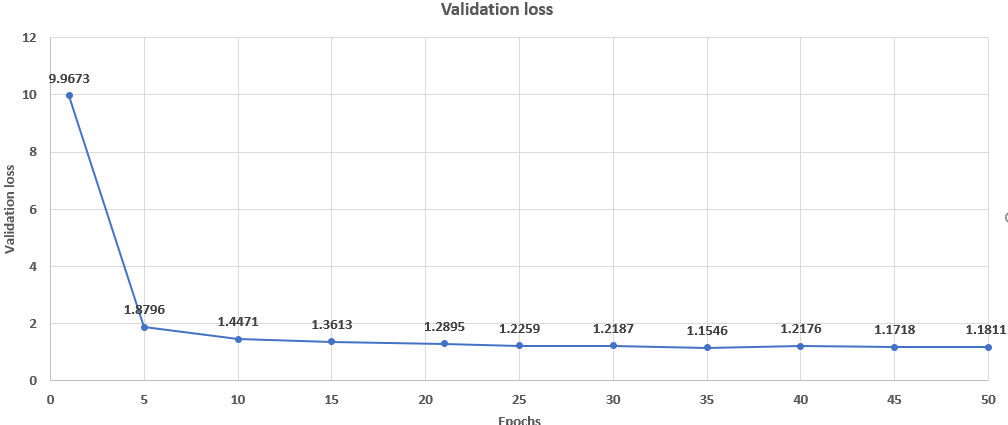}
\caption{Validation loss from epoch 1 to 50. The validation loss changes very slowly after 21 epochs and the change is not significant.}
\label{fig:min_loss}
\end{figure}
\begin{figure}[!t]
\centering
\graphicspath{{./figures/}}
\includegraphics[width=\columnwidth]{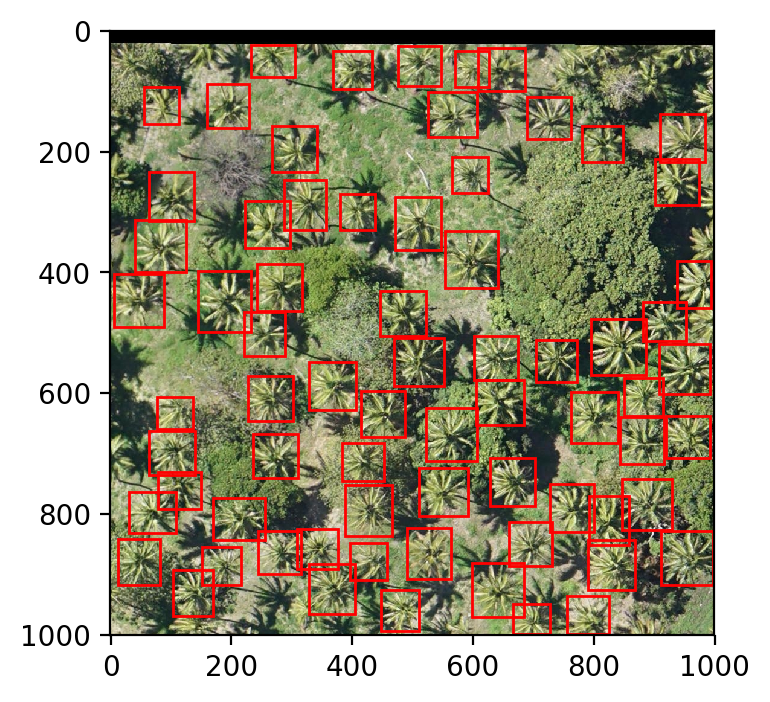}
\caption{Bounding box predictions using ResNet50. Many but not all coconut trees are detected. Few shadows are detected as coconut trees}
\label{fig:r1}
\end{figure}
\begin{figure}[!ht]
\centering
\graphicspath{{./figures/}}
\includegraphics[width=\columnwidth]{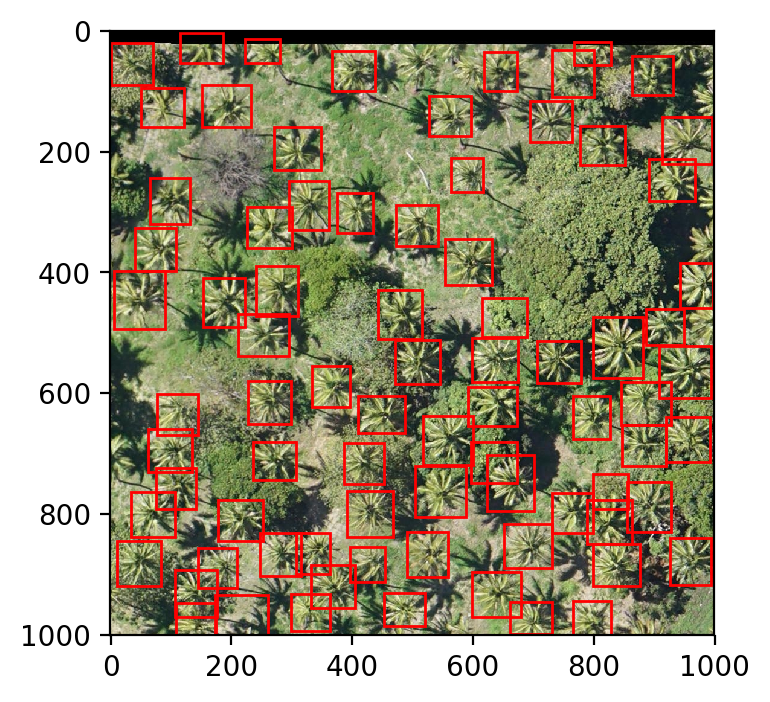}
\caption{Bounding box predictions using ResNet101. All coconut trees prediction confidences are above 70\%. The algorithm has incorrectly detected some coconut trees as threshold confidence is set low.}
\label{fig:Con_0.7}
\end{figure}
\begin{figure}[!ht]
\centering
\graphicspath{{./figures/}}
\includegraphics[width=\columnwidth]{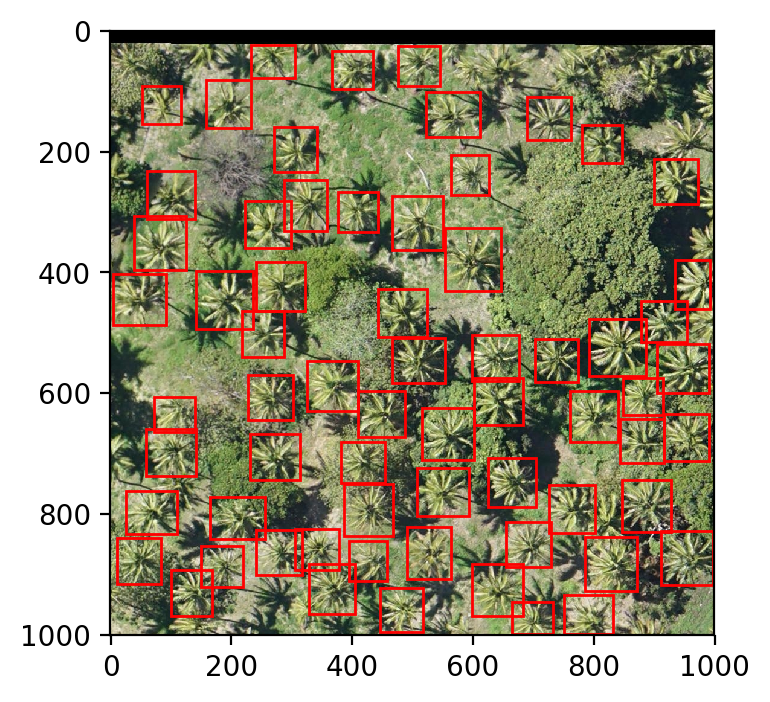}
\caption{Bounding box predictions using ResNet101. All coconut trees prediction confidences are above 95\%. The algorithm has missed out some coconut trees as threshold confidence is set high.}
\label{fig:Con_0.95}
\end{figure}
According to the validation scores of multiple experiments that we executed, we select the configuration having the best performance: train for 21 epochs, consist of 100 steps each, with the maximum number of ROIs being 110 (this is a good choice since we got a maximum of 70 trees per image tile as observed during the annotation phase).

\section{Results and Discussions}
\label{sec:results}
\begin{figure}[!t]
\centering
\graphicspath{{./figures/}}
\includegraphics[width=\columnwidth]{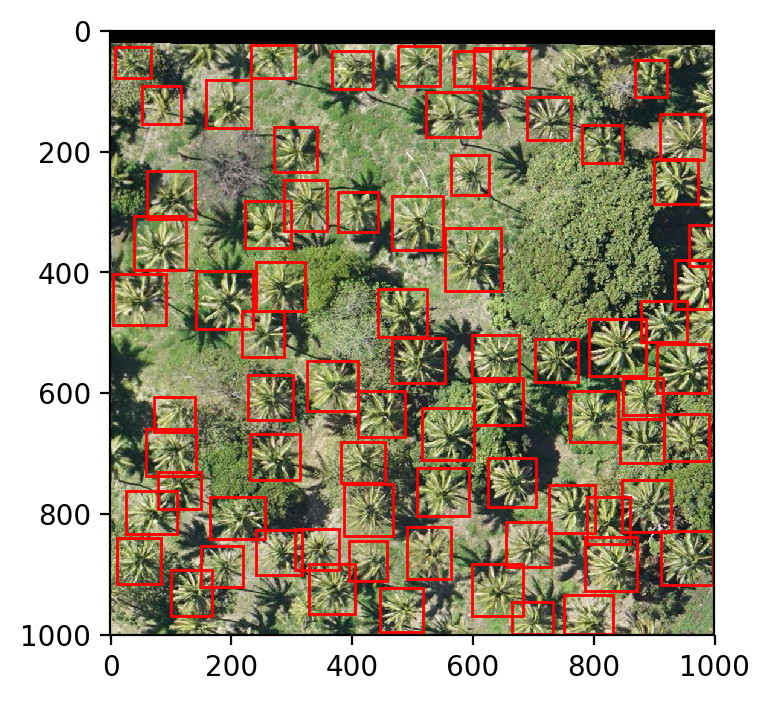}
\caption{Bounding box predictions using ResNet101. ResNet101 results are better than ResNet50 on the Bounding box predictions. All coconut trees prediction confidences are above 90\%}
\label{fig:r2}
\end{figure}

\begin{figure}[!ht]
\centering
\graphicspath{{./figures/}}
\includegraphics[width=\columnwidth]{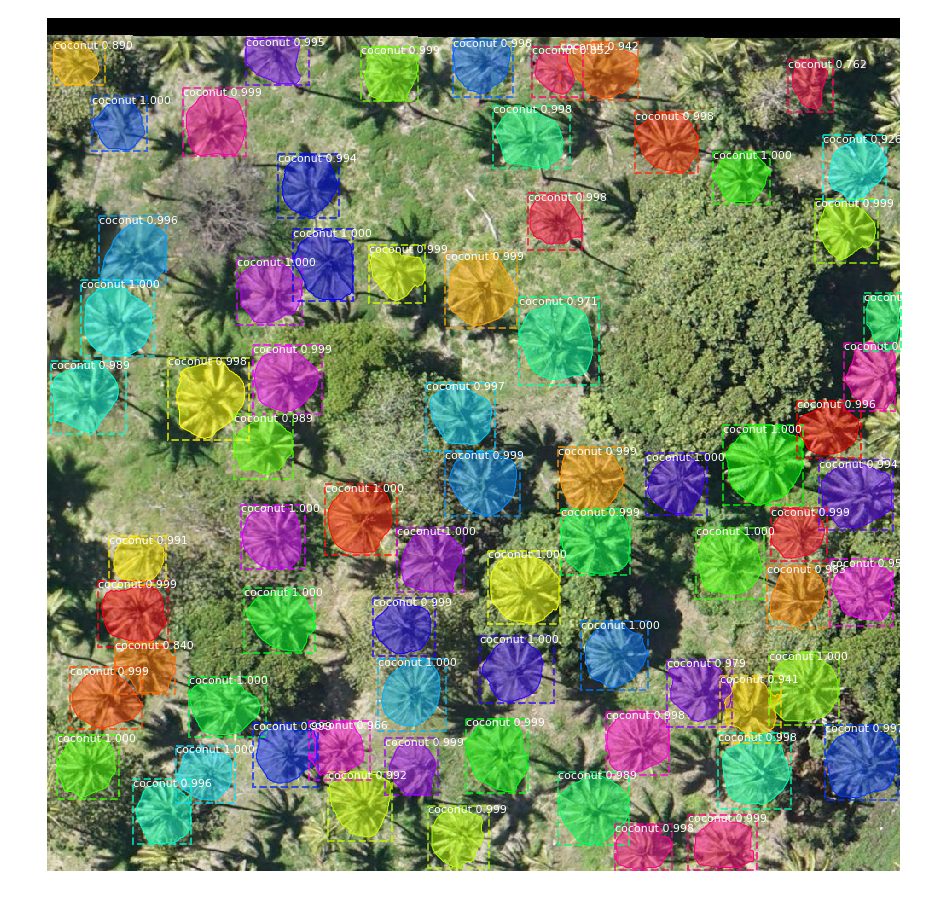}
\caption{Coconut trees prediction. Segmentation results using ResNet101 as backbone architecture. (Best seen in color)}
\label{fig:seg}
\end{figure}

We apply weights of our model trained for 21 epochs to detect coconut trees on the test-set that consist of 10 images\footnote{All the experiments are performed using Intel(R) Core i5-7300HQ CPU  2.50GHz (4 logical processors), with NVIDIA GeForce GTX 1050 4.00GB memory and 8GB of RAM on a 64-bit operating system.}. After performing several experiments best configuration settings are shown in Table 1. Our prediction results on some of the images are indicated in Fig. \ref{fig:r1} and Fig. \ref{fig:r2} The segmentation results are shown in Fig. \ref{fig:seg}

\begin{figure}[!ht]
\centering
\graphicspath{{./figures/}}
\includegraphics[width=\columnwidth]{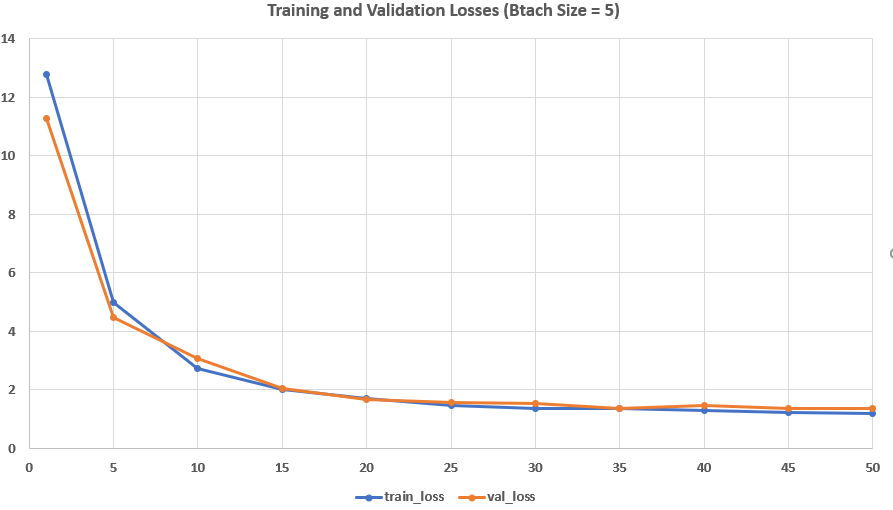}
\caption{Graph of train loss, validation loss of overall experiment reported up-till 50 epochs using batch size 5}
\label{fig:b1}
\end{figure}

\begin{figure}[!ht]
\centering
\graphicspath{{./figures/}}
\includegraphics[width=\columnwidth]{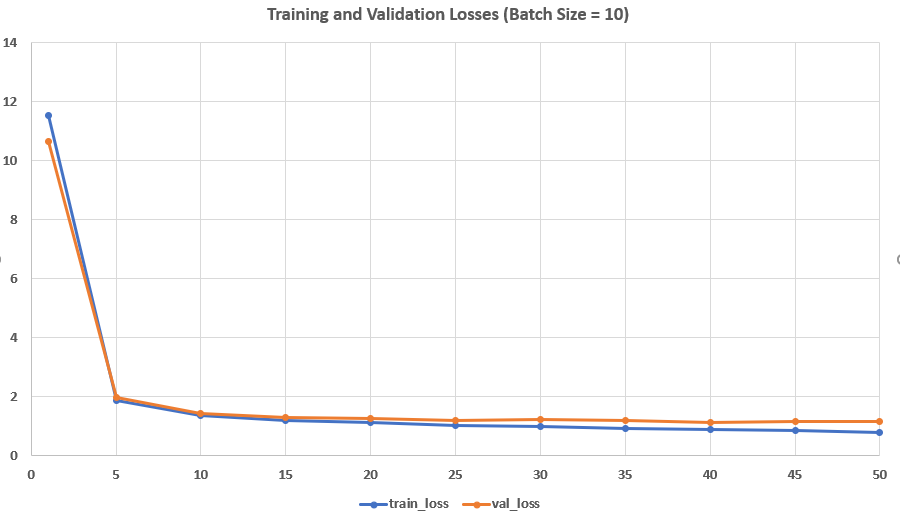}
\caption{Graph of train loss, validation loss of overall experiment reported up-till 50 epochs using batch size 10}
\label{fig:b2}
\end{figure}

The overall training time for 5 and 10 batch size was less compared to batch size 1 but training and validation loss of batch size of 5 and 10 is greater than batch size = 1. So, we select batch size = 1 for this experiment, as shown in Fig. \ref{fig:b1} and Fig. \ref{fig:b2}. A comparison table on training and validation loss for the different choices of batch sizes is shown in Table \ref{tab:tableBatchSizes}. 
Optimal learning rate of 0.001 (to avoid converging to local minima) after trying out different values among [0.0001, 0.0003, 0.001, 0.003, 0.01].

\subsection{Configuration Settings}
After performing several experiments, the best configuration settings which we set for our project are shown in Table I. 
\begin{table*}[!ht] 
\centering
\caption{Configuration settings\label{table:1}}
\begin{tabular}{|p{8cm}|p{2cm}|p{2cm}|} \hline
 \textbf{BACKBONE}&\textbf{ResNet101} \\\hline \rule{0pt}{10pt}
    BATCH SIZE & $1$  \\ \rule{0pt}{10pt}
    DETECTION MIN CONFIDENCE & $0.9$ \\ \rule{0pt}{8pt}
    DETECTION MAX INSTANCES & $100$ \\ \rule{0pt}{8pt}
    LEARNING MOMENTUM & $0.9$  \\ \rule{0pt}{8pt}
    LEARNING RATE & $0.001$ \\ \rule{0pt}{8pt}
    STEPS PER EPOCH & $100$  \\ \rule{0pt}{8pt}
    TRAIN ROIS PER IMAGE & $110$ \\ \rule{0pt}{8pt}
    VALIDATION STEPS & $50$  \\ \rule{0pt}{8pt}
    WEIGHT DECAY & $0.0001$ \\ \rule{0pt}{8pt}
    EPOCHS & $50$  \\ \hline
\end{tabular}
{}
\end{table*}
\subsection{Discussion}
\label{sec:Discussion}
\begin{figure}[!t]
\centering
\graphicspath{{./figures/}}
\includegraphics[width=\columnwidth]{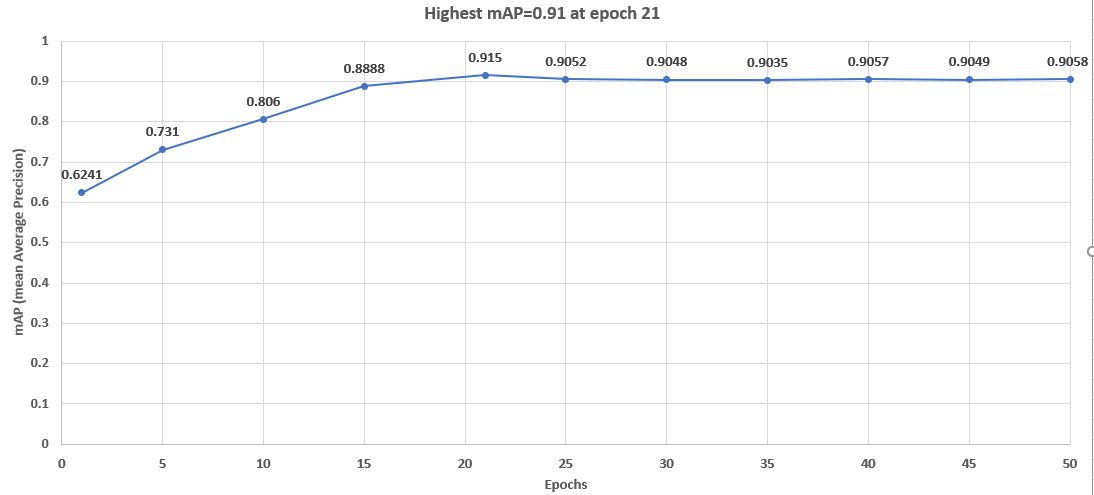}
\caption{Highest mean average precision (mAP) achieved is at epoch number 21 using ResNet101 as the backbone network in Mask R-CNN.}
\label{fig:map1}
\end{figure}

\begin{figure}[!t]
\centering
\graphicspath{{./figures/}}
\includegraphics[width=\columnwidth]{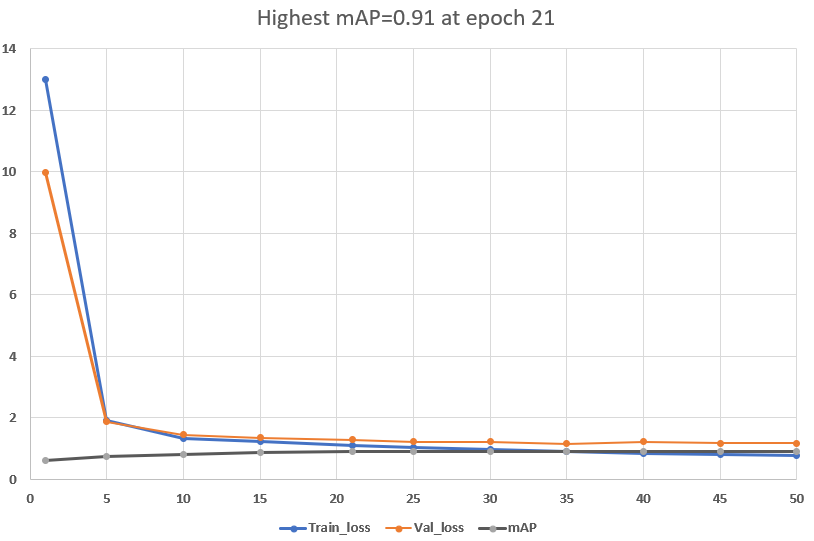}
\caption{Graph of train loss, validation loss and mean average precision (mAP) of overall experiment reported up-till 50 epochs. Highest mAP is recorded for epoch 21}
\label{fig:map2}
\end{figure}

We find that all coconut trees are detected with a considerably high confidence factor (which is $>90\%$). We achieved 96\% classification accuracy (CA) with ResNet50 and 98\% CA using ResNet101.  For a more formal evaluation, we have selected the mAP metric, a commonly used metric for performance evaluation of object detection. The mAP is a mean of average precision,  where not only the identification number but also the order of the correct predictions is evaluated. The highest mAP value achieved was 0.88 for ResNet50 and 0.91 for ResNet101. Detection results are visualized in Figure \ref{fig:r1} and Figure \ref{fig:r2}. The mAP curves are accordingly shown in Figure \ref{fig:map1} and Figure \ref{fig:map2}. F1 Score is 0.89 for ResNet50 and 0.92 for ResNet101. We have shown evaluation metrics in Table \ref{tab:table2}. The processing time of our approach is one minute for $1000\times1000$ image. Table \ref{tab:tableBatchSizes} summarizes the training and validation losses with different batch sizes used for training the model. The overall training and validation losses of batch size 1 and batch size 10 are almost similar despite batch size 10 having less training time. We have compared our proposed algorithm with some state-of-the-art techniques used for tree detection and segmentation, in the next sub-section.

\begin{table*}[!t]
\centering
\caption{Metrics of Model Performance\label{tab:table2}}
\begin{tabular}{llllll} \hline
 \textbf{Model}&\textbf{Mean Average}&\textbf{Precision}&\textbf{Recall}&\textbf{F1}&\textbf{Classification} \\
\textbf{}&\textbf{Precision}&\textbf{}&\textbf{}&\textbf{ Score}&\textbf{Accuracy} \\\hline \rule{0pt}{10pt}
    ResNet50 & $88\%$  & $91\%$ & $85\%$& $89\%$ & $96\%$\\ \rule{0pt}{10pt}
    ResNet101 & $91\%$  & $96.9\%$ & $88\%$& $92\%$ & $98\%$\\ \hline
\end{tabular}
{}
\end{table*}
\begin{table}[!t]
\centering
\caption{Comparison of batch sizes against training and validation losses\label{tab:tableBatchSizes}}

\begin{tabular}{llll} \hline
 \textbf{Batch Size}&\textbf{Epochs} & \textbf{Train loss} & \textbf{Validation loss} \\
 \hline \rule{0pt}{10pt}
    1 & 1 & 13.014 & 9.9673 \\ \hline \rule{0pt}{10pt}
    & 5 & 1.9152 & 1.8796 \\ \hline \rule{0pt}{10pt}
 & 10 & 1.3307 & 1.4471 \\ \hline \rule{0pt}{10pt}
 & 15 & 1.216 & 1.3613 \\ \hline \rule{0pt}{10pt}
 & 20 & 1.0977 & 1.2895 \\ \hline \rule{0pt}{10pt}
 & 25 & 1.0229 & 1.2259 \\ \hline \rule{0pt}{10pt}
 & 30 & 0.9644 & 1.2187 \\ \hline \rule{0pt}{10pt}
 & 35 & 0.9103 & 1.1546 \\ \hline \rule{0pt}{10pt}
 & 40 & 0.8516 & 1.2176 \\ \hline \rule{0pt}{10pt}
 & 45 & 0.8219 & 1.1718 \\ \hline \rule{0pt}{10pt}
 & 50 & 0.7823 & 1.1811 \\ \hline \rule{0pt}{10pt}
 \textbf{Batch Size} & \textbf{Epochs} & \textbf{Train Loss} & \textbf{Val Loss} \\ \hline \rule{0pt}{10pt}
5 & 1 & 12.775 & 11.29 \\ \hline \rule{0pt}{10pt}
 & 5 & 4.9746 & 4.469 \\ \hline \rule{0pt}{10pt}
 & 10 & 2.7309 & 3.0899 \\ \hline \rule{0pt}{10pt}
 & 15 & 1.9974 & 2.0535 \\ \hline \rule{0pt}{10pt}
 & 20 & 1.6909 & 1.6614 \\ \hline \rule{0pt}{10pt}
 & 25 & 1.4612 & 1.5781 \\ \hline \rule{0pt}{10pt}
 & 30 & 1.3665 & 1.5517 \\ \hline \rule{0pt}{10pt}
 & 35 & 1.3693 & 1.3778 \\ \hline \rule{0pt}{10pt}
 & 40 & 1.2955 & 1.4525 \\ \hline \rule{0pt}{10pt}
 & 45 & 1.2143 & 1.3676 \\ \hline \rule{0pt}{10pt}
 & 50 & 1.1865 & 1.3668 \\ \hline \rule{0pt}{10pt}
\textbf{Batch Size} & \textbf{Epochs} & \textbf{Train Loss} & \textbf{Val Loss} \\ \hline \rule{0pt}{10pt}
10 & 1 & 11.519 & 10.633 \\ \hline \rule{0pt}{10pt}
 & 5 & 1.8724 & 1.9503 \\ \hline \rule{0pt}{10pt}
 & 10 & 1.3511 & 1.4069 \\ \hline \rule{0pt}{10pt}
 & 15 & 1.1707 & 1.2847 \\ \hline \rule{0pt}{10pt}
 & 20 & 1.1045 & 1.2601 \\ \hline \rule{0pt}{10pt}
 & 25 & 1.0186 & 1.1916 \\ \hline \rule{0pt}{10pt}
 & 30 & 0.969 & 1.2302 \\ \hline \rule{0pt}{10pt}
 & 35 & 0.9195 & 1.1932 \\ \hline \rule{0pt}{10pt}
 & 40 & 0.8686 & 1.1225 \\ \hline \rule{0pt}{10pt}
 & 45 & 0.8296 & 1.1439 \\ \hline \rule{0pt}{10pt}
 & 50 & 0.7795 & 1.1556 \\ \hline \rule{0pt}{10pt}
\end{tabular}
{}
\end{table}

\begin{table*}[t]
\centering
\caption{Comparison with other techniques\label{tab:table3}}

\begin{tabular}{|p{3cm}|p{3cm}|p{3.5cm}|p{3.5cm}|}
\hline
\textbf{Reference}&\textbf{Data}&\textbf{Model}&\textbf{Score}\\
 \hline \rule{0pt}{10pt}

Chen et al. $(2014)$ \cite{reff19chen2015spectral} & Hyperspectral imagery  & Hybrid of PCA, logistic regression, and autoencoder & Pavia: 84.62\%(CA) 0.8451\% (F1) \\\hline \rule{0pt}{10pt}
  
Luus et al. $(2015) $\cite{reff30luus2015multivie},& Aerial imagery & Author defined CNN & 93.48\% (CA)  \\\hline \rule{0pt}{10pt}

Mortensen et al. $(2016)$ \cite{reff33mortensen2016semantic},& Photograph by sonny a7  & Adapted version of VGG16 & 79\%(CA), 0.66\% (IOU)\\\hline \rule{0pt}{10pt}

Lu et al. $(2017)$ \cite{reff34lu2017cultivated},& UAV imagery  & Author defined CNN & 89.5\% (CA)\\\hline \rule{0pt}{10pt}

Milioto et al. $(2017) $\cite{reff29milioto2017real},& UAV imagery  & Author defined CNN & 97.50\% (CA)\\\hline \rule{0pt}{10pt}

Sorensen et al. $(2017)$ \cite{reff31jonquet2017agroportal},& Photograph by Canon PowerShot G15  & DenseNet & 97.0\% (CA)\\\hline \rule{0pt}{10pt}

Saldana et al. $(2019)$ \cite{reff32ochoa2019framework},& Aerial imagery  & Adapted version of SegNet and YOLO & 97.5\% (CA),0.89 (F1)\\\hline \rule{0pt}{10pt}

Ours, & UAV aerial imagery  & Adapted version of Mask RCNN
With backbone ResNet50 & 96\% (CA), 89\% (F1), 88\% (mAP) \\\hline \rule{0pt}{10pt}

Ours, & UAV aerial imagery  & Adapted version of Mask RCNN
With backbone ResNet101  & 98\%(CA), 92\% (F1), 91\% (mAP)\\ \hline

\end{tabular}
{}
\end{table*}

\subsection{Comparison with other techniques}
Some excellent works reported on trees detection from high-resolution aerial imagery involve different datasets, pre-processing methods, models, parameters, and metrics. We do not make a direct comparison as the datasets used or the tasks performed in these approaches are different.  However, it is still useful to provide a summary of the results of these approaches \cite{reff28kamilaris2018deep}. A summary of the data, model, score, and performance is reported in Table \ref{tab:table3}.
As discussed earlier the performance score of the algorithm differs depending upon the task. So, we have compared our results considering the ones that have used a similar performance. Milioto et al., \cite{reff29milioto2017real} have reported accuracy of 84.62\% on classification tasks while using a hybrid of PCA, logistic regression, and autoencoder. Luus et al., \cite{reff30luus2015multivie} have reported an accuracy of 93.48\% on classification task while using CNNs. Sorensen et al.\cite{reff31jonquet2017agroportal} has reported an accuracy of 97\% on classification task  while using DenseNet. Saldana et al., \cite{reff32ochoa2019framework} has reported an 80\% localization accuracy, 97.5\% classification accuracy, and 0.89 of F1 score on localization and segmentation task while using an adapted version of YOLO and SegNet. In our work, we achieved 96\% classification accuracy with ResNet50 and 98\% classification accuracy achieved with ResNet101. F1 score of resnet50 is 89\% and using ResNet101 it is 92\%. It is worth mentioning that all of the aforementioned experiments except (Saldana et al., \cite{reff32ochoa2019framework}) dealt only with the classification task. We propose an approach which not only performs classification but also locate coconut trees and segment the trees. We evaluate an additional performance metric, the mean Average Precision (mAP). We achieved 88\% mAP using ResNet50 and 91\% mAP with backbone architecture ResNet101. This metric shows how accurate the model is to locate and classify the coconut trees. 

\section{Conclusion and Future Direction}
\label{sec:Conclusion}
In this paper, we have presented an approach for coconut trees detection and segmentation in aerial imagery of the kingdom of Tonga (South Pacific Islands). We have reported a Mask R-CNN based model using ResNet50 and ResNet101 backbone architectures. The model is trained on the data which we processed and prepared from a single high-resolution aerial image along with the shape file. Experimental results have shown that our model is able to predict coconut trees with quite a high accuracy (91\% mean average precision). Our model can be effortlessly extended to classify and locate other kinds of food trees as well. A comparative setup showed that we get better accuracy for the ResNet101 architecture when compared with the performance of a ResNet50 based model. Moreover, it carries the benefits of faster R-CNN which is faster than conventional R-CNN and more accurate than CNN. The work carries significance in food resource assessment, humanitarian aid services, and damage analysis in disaster-hit areas, using high-resolution satellite imagery.

The research work is one of the attempts to classify and locate coconut trees based on remote sensed aerial imagery dataset. There is much more potential for future studies in this area. One task of particular significance is to get a cleaner dataset and have methods to get better annotations as these will improve the model training. Our future task includes model development to detect other types of food trees (mango, banana, papaya), as well as road conditions and their types.

\bibliographystyle{IEEEtran} %
\bibliography{paper}

\begin{thebibliography}{10}
\providecommand{\url}[1]{#1}
\csname url@samestyle\endcsname
\providecommand{\newblock}{\relax}
\providecommand{\bibinfo}[2]{#2}
\providecommand{\BIBentrySTDinterwordspacing}{\spaceskip=0pt\relax}
\providecommand{\BIBentryALTinterwordstretchfactor}{4}
\providecommand{\BIBentryALTinterwordspacing}{\spaceskip=\fontdimen2\font plus
\BIBentryALTinterwordstretchfactor\fontdimen3\font minus
  \fontdimen4\font\relax}
\providecommand{\BIBforeignlanguage}[2]{{%
\expandafter\ifx\csname l@#1\endcsname\relax
\typeout{** WARNING: IEEEtran.bst: No hyphenation pattern has been}%
\typeout{** loaded for the language `#1'. Using the pattern for}%
\typeout{** the default language instead.}%
\else
\language=\csname l@#1\endcsname
\fi
#2}}
\providecommand{\BIBdecl}{\relax}
\BIBdecl

\bibitem{reff1}
A.~Fritz, ``Tropical cyclone gita is a monster category 4, and it's hammering
  tonga,'' in \emph{The Washington Post}.\hskip 1em plus 0.5em minus
  0.4em\relax The Washington Post, [Online]. Available:
  https://www.washingtonpost.com/ [Accessed: 14-May-2019, 2018.

\bibitem{reff2}
U.~N.-G. Assembly, ``International strategy for disaster reduction,'' in
  \emph{A/RES/70/204}.\hskip 1em plus 0.5em minus 0.4em\relax United
  Nations-General Assembly, 2015.

\bibitem{reff3}
K.~Leetaru, ``Using ai for good: A new data challenge to use ai to triage
  natural disaster aerial imagery,'' in \emph{Forbes}.\hskip 1em plus 0.5em
  minus 0.4em\relax Forbes [Online]. Available: https://www.forbes.com
  [Accessed: 14-May-2019], 2018.

\bibitem{reff4}
W.~Bank, ``World bank: Automated feature detection of aerial imagery from south
  pacific - live - google docs.''\hskip 1em plus 0.5em minus 0.4em\relax Google
  Docs [Online]. [Accessed: 14-May-2019, 2018.

\bibitem{reff5yang2018deep}
M.~Y. Yang, W.~Liao, X.~Li, and B.~Rosenhahn, ``Deep learning for vehicle
  detection in aerial images,'' in \emph{2018 25th IEEE International
  Conference on Image Processing (ICIP)}.\hskip 1em plus 0.5em minus
  0.4em\relax IEEE, 2018, pp. 3079--3083.

\bibitem{reff6sommer2017fast}
L.~W. Sommer, T.~Schuchert, and J.~Beyerer, ``Fast deep vehicle detection in
  aerial images,'' in \emph{{2017 IEEE Winter Conference on Applications of
  Computer Vision (WACV)}}.\hskip 1em plus 0.5em minus 0.4em\relax IEEE, 2017,
  pp. 311--319.

\bibitem{reff7luque2017spatio}
B.~Luque, J.~R. Morros~Rubi{\'o}, and J.~Ruiz~Hidalgo, ``Spatio-temporal road
  detection from aerial imagery using cnns,'' in \emph{Proceedings of the 12th
  International Joint Conference on Computer Vision, Imaging and Computer
  Graphics Theory and Applications, Volume 4: VISAPP}.\hskip 1em plus 0.5em
  minus 0.4em\relax SCITEPRESS, 2017, pp. 493--500.

\bibitem{reff8dalponte2008fusion}
M.~Dalponte, L.~Bruzzone, and D.~Gianelle, ``Fusion of hyperspectral and lidar
  remote sensing data for classification of complex forest areas,'' \emph{IEEE
  Transactions on Geoscience and Remote Sensing}, vol.~46, no.~5, pp.
  1416--1427, 2008.

\bibitem{reff9frolking2002combining}
S.~Frolking, J.~Qiu, S.~Boles, X.~Xiao, J.~Liu, Y.~Zhuang, C.~Li, and X.~Qin,
  ``Combining remote sensing and ground census data to develop new maps of the
  distribution of rice agriculture in china,'' \emph{Global Biogeochemical
  Cycles}, vol.~16, no.~4, pp. 38--1, 2002.

\bibitem{reff10rhee2010monitoring}
J.~Rhee, J.~Im, and G.~J. Carbone, ``Monitoring agricultural drought for arid
  and humid regions using multi-sensor remote sensing data,'' \emph{Remote
  Sensing of Environment}, vol. 114, no.~12, pp. 2875--2887, 2010.

\bibitem{reff11li2010saliency}
Z.~Li and L.~Itti, ``Saliency and gist features for target detection in
  satellite images,'' \emph{IEEE Transactions on Image Processing}, vol.~20,
  no.~7, pp. 2017--2029, 2010.

\bibitem{reff12campbell2011introduction}
J.~B. Campbell and R.~H. Wynne, \emph{Introduction to remote sensing}.\hskip
  1em plus 0.5em minus 0.4em\relax Guilford Press, 2011.

\bibitem{reff13schmidhuber2015deep}
J.~Schmidhuber, ``Deep learning in neural networks: An overview,'' \emph{Neural
  networks}, vol.~61, pp. 85--117, 2015.

\bibitem{reff14rusk2015deep}
N.~Rusk, ``Deep learning,'' \emph{Nature Methods}, vol.~13, no.~1, p.~35, 2015.

\bibitem{reff15pan2009survey}
S.~J. Pan and Q.~Yang, ``A survey on transfer learning,'' \emph{IEEE
  Transactions on knowledge and data engineering}, vol.~22, no.~10, pp.
  1345--1359, 2009.

\bibitem{reff16demmers2012simultaneous}
T.~G. Demmers, Y.~Cao, D.~J. Parsons, S.~Gauss, and C.~M. Wathes,
  ``Simultaneous monitoring and control of pig growth and ammonia emissions,''
  in \emph{2012 IX International Livestock Environment Symposium (ILES
  IX)}.\hskip 1em plus 0.5em minus 0.4em\relax American Society of Agricultural
  and Biological Engineers, 2012, p.~3.

\bibitem{reff17sehgal2017crop}
G.~Sehgal, B.~Gupta, K.~Paneri, K.~Singh, G.~Sharma, and G.~Shroff, ``Crop
  planning using stochastic visual optimization,'' in \emph{2017 IEEE
  Visualization in Data Science (VDS)}.\hskip 1em plus 0.5em minus 0.4em\relax
  IEEE, 2017, pp. 47--51.

\bibitem{reff18amara2017deep}
J.~Amara, B.~Bouaziz, A.~Algergawy \emph{et~al.}, ``A deep learning-based
  approach for banana leaf diseases classification.'' in \emph{BTW
  (Workshops)}, 2017, pp. 79--88.

\bibitem{reff20zhao2019object}
Z.-Q. Zhao, P.~Zheng, S.-t. Xu, and X.~Wu, ``Object detection with deep
  learning: A review,'' \emph{IEEE transactions on neural networks and learning
  systems}, 2019.

\bibitem{reff21girshick2015region}
R.~Girshick, J.~Donahue, T.~Darrell, and J.~Malik, ``Region-based convolutional
  networks for accurate object detection and segmentation,'' \emph{IEEE
  transactions on pattern analysis and machine intelligence}, vol.~38, no.~1,
  pp. 142--158, 2015.

\bibitem{reff22girshick2015fast}
R.~Girshick, ``Fast r-cnn,'' in \emph{Proceedings of the IEEE international
  conference on computer vision}, 2015, pp. 1440--1448.

\bibitem{reff23ren2015faster}
S.~Ren, K.~He, R.~Girshick, and J.~Sun, ``Faster r-cnn: Towards real-time
  object detection with region proposal networks,'' in \emph{Advances in neural
  information processing systems}, 2015, pp. 91--99.

\bibitem{reff24he2017mask}
K.~He, G.~Gkioxari, P.~Doll{\'a}r, and R.~Girshick, ``Mask r-cnn,'' in
  \emph{Proceedings of the IEEE international conference on computer vision},
  2017, pp. 2961--2969.

\bibitem{maghsoudi2020automatic}
O.~H. Maghsoudi, A.~Gastounioti, L.~Pantalone, E.~Conant, and D.~Kontos,
  ``Automatic breast segmentation in digital mammography using a convolutional
  neural network,'' in \emph{15th International Workshop on Breast Imaging
  (IWBI2020)}, vol. 11513.\hskip 1em plus 0.5em minus 0.4em\relax International
  Society for Optics and Photonics, 2020, p. 1151322.

\bibitem{liang2020polytransform}
J.~Liang, N.~Homayounfar, W.-C. Ma, Y.~Xiong, R.~Hu, and R.~Urtasun,
  ``Polytransform: Deep polygon transformer for instance segmentation,'' in
  \emph{Proceedings of the IEEE/CVF Conference on Computer Vision and Pattern
  Recognition}, 2020, pp. 9131--9140.

\bibitem{richardsocher2017}
N.~S. Keskar and R.~Socher, ``Improving generalization performance by switching
  from adam to sgd,'' \emph{ArXiv}, vol. abs/1712.07628, 2017.

\bibitem{reff27he2016deep}
K.~He, X.~Zhang, S.~Ren, and J.~Sun, ``Deep residual learning for image
  recognition,'' in \emph{Proceedings of the IEEE conference on computer vision
  and pattern recognition}, 2016, pp. 770--778.

\bibitem{reff19chen2015spectral}
Y.~Chen, X.~Zhao, and X.~Jia, ``Spectral--spatial classification of
  hyperspectral data based on deep belief network,'' \emph{IEEE Journal of
  Selected Topics in Applied Earth Observations and Remote Sensing}, vol.~8,
  no.~6, pp. 2381--2392, 2015.

\bibitem{reff30luus2015multivie}
F.~P. Luus, B.~P. Salmon, F.~Van~den Bergh, and B.~T.~J. Maharaj, ``Multiview
  deep learning for land-use classification,'' \emph{IEEE Geoscience and Remote
  Sensing Letters}, vol.~12, no.~12, pp. 2448--2452, 2015.

\bibitem{reff33mortensen2016semantic}
A.~K. Mortensen, M.~Dyrmann, H.~Karstoft, R.~N. J{\o}rgensen, R.~Gislum
  \emph{et~al.}, ``Semantic segmentation of mixed crops using deep
  convolutional neural network.'' in \emph{CIGR-AgEng Conference, 26-29 June
  2016, Aarhus, Denmark. Abstracts and Full papers}.\hskip 1em plus 0.5em minus
  0.4em\relax Organising Committee, CIGR 2016, 2016, pp. 1--6.

\bibitem{reff34lu2017cultivated}
H.~Lu, X.~Fu, C.~Liu, L.-g. Li, Y.-x. He, and N.-w. Li, ``Cultivated land
  information extraction in uav imagery based on deep convolutional neural
  network and transfer learning,'' \emph{Journal of Mountain Science}, vol.~14,
  no.~4, pp. 731--741, 2017.

\bibitem{reff29milioto2017real}
A.~Milioto, P.~Lottes, and C.~Stachniss, ``Real-time blob-wise sugar beets vs
  weeds classification for monitoring fields using convolutional neural
  networks,'' \emph{ISPRS Annals of the Photogrammetry, Remote Sensing and
  Spatial Information Sciences}, vol.~4, p.~41, 2017.

\bibitem{reff31jonquet2017agroportal}
C.~Jonquet, A.~Toulet, E.~Arnaud, S.~Aubin, E.~D. Yeumo, V.~Emonet,
  J.~Graybeal, M.-A. Laporte, M.~A. Musen, V.~Pesce \emph{et~al.},
  ``Agroportal: A vocabulary and ontology repository for agronomy,''
  \emph{Computers and Electronics in Agriculture}, vol. 144, pp. 126--143,
  2018.

\bibitem{reff32ochoa2019framework}
K.~S. Ochoa and Z.~Guo, ``A framework for the management of agricultural
  resources with automated aerial imagery detection,'' \emph{Computers and
  Electronics in Agriculture}, vol. 162, pp. 53--69, 2019.

\bibitem{reff28kamilaris2018deep}
A.~Kamilaris and F.~X. Prenafeta-Bold{\'u}, ``Deep learning in agriculture: A
  survey,'' \emph{Computers and electronics in agriculture}, vol. 147, pp.
  70--90, 2018.

\end{thebibliography}

\end{document}